# Reliability of AI/ML Computational Searches for Magnetic Materials: Databases, Validation, and Synthesizability

Vladimir Antropov[1,3], Subhadip Pradhan[2], Kirill D. Belashchenko[2]

[1]Department of Physics and Astronomy, Iowa State University, Ames, IA 50011, USA
[2]Department of Physics and Astronomy and Nebraska Center for Materials and Nanoscience, University of Nebraska-Lincoln, Lincoln, NE 68588, USA
[3]Ames National Laboratory, U.S. Department of Energy, Ames, IA 50011, USA

**The rapid growth of AI/ML methods for materials discovery has increased the importance of the quality, physical completeness, and information content of computational databases. We examine these issues for magnetic materials, where incomplete sampling of competing magnetic states, chemical disorder, and finite-temperature properties can lead to apparently accurate but physically unreliable predictions. We propose a hierarchical discovery workflow combining independent ML models with explicit electronic-structure, magnetic, dynamical, thermodynamic, kinetic, and experimental validation. To incorporate kinetic accessibility into high-throughput screening, we introduce a dimensionless synthesis-window descriptor based on the relation between the chemical order-disorder and Tammann temperatures. A cluster-expansion analysis of Fe-Co-B illustrates the importance of this criterion: although ordered $Fe_3CoB_2$ is energetically favored at T=0, its small ordering energy produces an order-disorder temperature far below room temperature, making long-range Fe/Co ordering kinetically unachievable. These results demonstrate that reliable AI/ML discovery of magnetic materials requires databases and validation procedures that contain physical information relevant to the target functionality, rather than simply larger numbers of calculated structures.**

## I. Introduction

The rapid proliferation of artificial intelligence (AI) and machine learning (ML) methods in materials science recalls the phenomenon Claude E. Shannon discussed in his 1956 editorial entitled The Bandwagon [1], which remains remarkably relevant to today's explosion of AI/ML. After replacing the words *"information theory"* with *"artificial intelligence"* or *"machine learning,"* much of Shannon's warning reads as if it were written for today. Shannon cautioned that a powerful new methodology's success can spur its application beyond the domain where its reliability has been established. This warning is particularly relevant to contemporary AI/ML-based materials discovery, where increasingly sophisticated algorithms are often applied to rapidly expanding computational databases under the implicit assumption that database size measures the amount of information available for learning.

From an information-theoretic perspective, however, data volume and information content are fundamentally different quantities. For a distribution of materials among physically relevant classes ($i$), a simple measure of diversity can be expressed through the Shannon entropy, $S_{inf} = -\sum_i p_i \ln p_i$ , where $p_i$ denotes the probability of sampling class ($i$). A database can therefore increase substantially in size while acquiring relatively little new information if additional entries are redundant or concentrated within already well-sampled regions of chemical and structural space. Conversely, information about the missing classes of structures, electronic states, or magnetic configurations cannot, in general, be reconstructed simply by increasing an ML model's complexity. The missing classes may represent structural prototypes, chemical families, magnetic configurations, or other physically relevant partitions; the appropriate data partitioning depends on the target property, so the entropy is used here as an illustrative measure of database diversity rather than a universal metric.

This distinction is particularly important for magnetism. Experimental databases record realized materials, whereas computational databases contain model-generated states whose physical relevance depends on the approximations and sampling protocols used to construct them. General-purpose high-throughput DFT databases were not designed as comprehensive databases of magnetic ground states, and their magnetic information is therefore incomplete and nonuniform. Reproducibility of computational entries also depends on the full electronic-structure protocol, including magnetic initialization and the treatment of correlations and spin-orbit coupling, not only on the exchange-correlation functional. For example, comparison of metallic Fe-Co-based systems in the ML-generated Alexandria database with direct DFT searches using USPEX or genetic-algorithm methods reveals structural and magnetic disagreements in predicted ground states.

A practical example illustrates why database completeness must include the degrees of freedom relevant to the target property. We performed a computational DFT search for new rare-earth-free magnets using the Alexandria database, imposing three constraints: magnetic anisotropy > 1 MJ/m3, Curie temperature > 600 K, and magnetization > 1 T. Twelve ordered systems were found with parameters close to the required values. However, after including chemical disorder, DFT eliminated all newly identified ordered systems. This result emphasizes the need for magnetism-oriented computational databases that contain not only crystal structures and T=0 energies but also competing magnetic states, chemical disorder, and, where relevant to the target application, finite-temperature magnetic information. Constructing such magnetic content solely through large numbers of unconstrained self-consistent calculations is inefficient; established magnetism methods can often determine likely magnetic states more directly.

The consequences extend directly to AI/ML predictions. Different ML architectures, descriptors, training subsets, optimization procedures, and sampling strategies can produce substantially different predictions for the same unexplored materials. Such disagreement is usually treated as undesirable model uncertainty, but it can also be informative: if independently constructed models reproduce the available training and validation data with comparable accuracy yet disagree in a particular region of materials space, their divergence empirically indicates that the available training information poorly constrains that region.

Model-to-model disagreement may therefore serve as a practical measure of an information deficit in the underlying database. For an observable (X) predicted by (M) models, a simple measure is the model variance

$$\sigma_{ML}^2(X) = \frac{1}{M}\sum_{\alpha=1}^{M}[X_\alpha - \bar{X}]^2$$

where $X_\alpha$ is the prediction of model $\alpha$ and $\bar{X}$ is their mean. A large $\sigma_{ML}^2$ does not constitute a Shannon entropy by itself, since model disagreement can also originate from differences in algorithms, descriptors, or optimization. Nevertheless, after controlling these effects, increased intermodel dispersion can serve as an operational indicator of insufficient training information. In this sense, disagreement among AI/ML predictions is not merely a weakness of the methodology but potentially a diagnostic of poorly sampled regions of materials space.

The problem is especially acute for magnetic materials because crystallographic structure does not uniquely determine its magnetic ground state. Missing magnetic configurations, incomplete sampling of competing phases, inconsistent magnetic initialization, missing metallic systems, and approximations in the underlying electronic-structure calculations can introduce systematic biases that AI/ML models subsequently learn. Consequently, a large number of calculated structures does not necessarily provide a comparably large amount of independent magnetic information, and agreement with a randomly selected test set does not guarantee reliability when a model is applied to a new chemical, structural, or magnetic class.

The database must also contain the physics required by the target application. For permanent magnets, room-temperature magnetic anisotropy is essential: a search restricted to T=0 K anisotropy would fail to identify even $Nd_2Fe_{14}B$ appropriately as a permanent magnet because its anisotropy changes strongly with temperature. Useful permanent-magnet databases should aim to represent at least intrinsic magnetic properties at relevant temperatures, rather than relying only on T=0 total energies.

The AI/ML methods should therefore be assessed not only by model and database size but by the independent, physically relevant information contained in the data. In the spirit of Shannon's warning, the central question is not simply how many entries are available, but whether the database samples the structural, chemical, magnetic, and finite-temperature degrees of freedom needed for the intended prediction. Model-to-model variance can complement these database-based information measures by identifying regions where predictions are weakly constrained. Otherwise, we will always deal with a well-known principle (see Fig.1).

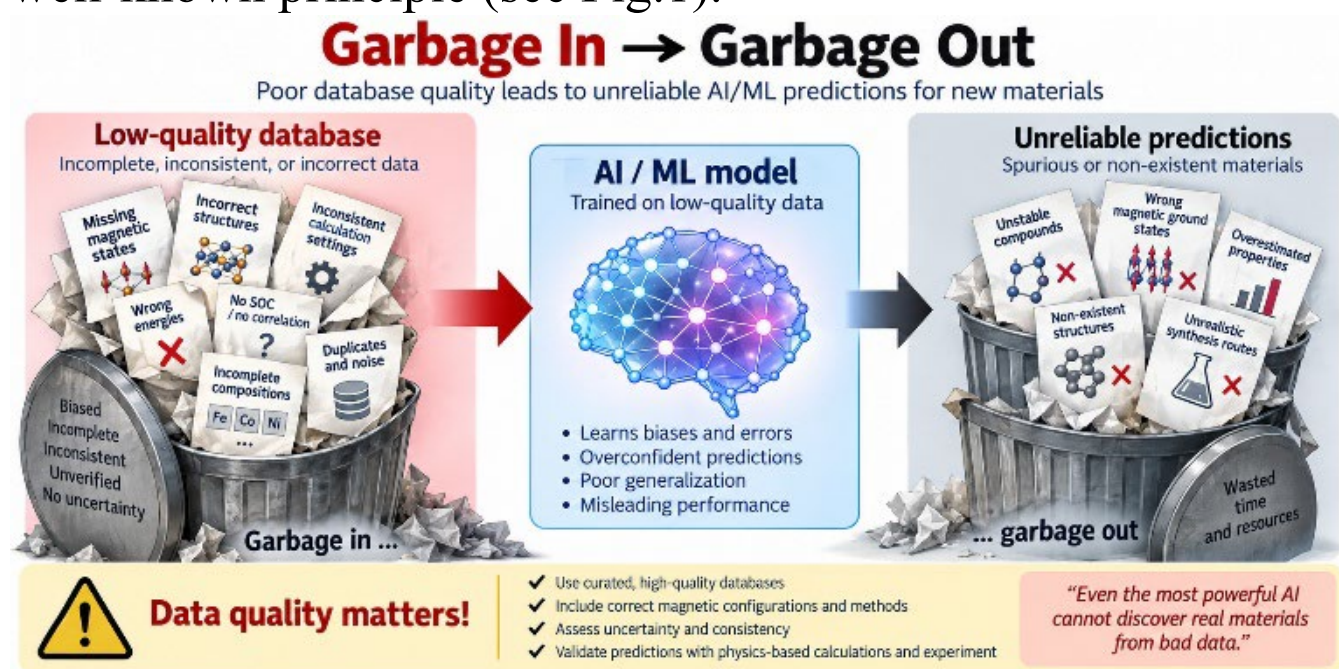

Fig.1 Illustration of the importance of reliable databases.

## II. Usage of AI/ML algorithms

The preceding considerations suggest the hierarchical workflow shown in Fig. 2, in which each stage supplies physical information that is generally absent from the preceding stage. The first step uses several ML models with different representations or architectures to generate and prioritize a broad candidate space. Agreement among independent models provides an initial robustness check, while disagreement can identify regions that deserve additional sampling or direct calculation. A useful benchmark is Matbench Discovery [2], which evaluates ML models prospectively on structures outside their training data and reconstructs hull distances from predicted formation energies. Although newer models achieve formation-energy errors of roughly 18-25 meV/atom [2], these values should not be interpreted as the uncertainty of the hull position of an individual new material; good regression performance does not necessarily imply reliable stable/unstable classification near the convex hull. For magnetic materials, the challenge is greater because the relevant magnetic energy scales can be smaller, and competing magnetic configurations are often incompletely represented in the training data.

Current ML algorithms for magnetic systems [3] report prediction errors of around 20 meV. This energy scale can exclude potentially relevant magnetic candidates and sequentially applying structural-stability and magnetic ML models can compound their uncertainties. We therefore advocate for using ML primarily to generate a broad stability-based candidate space, followed by targeted high-throughput DFT calculations that explicitly examine competing magnetic states.

Model consensus is not sufficient because models trained on the same database can inherit common systematic errors. Explicit

DFT calculations are therefore needed to verify crystal structure, formation energy, magnetic state, and stability related to competing phases. For rare-earth magnets, the current absence of a uniformly reliable structural and magnetic database makes this validation especially important.

The subsequent filters test physically distinct forms of stability and accessibility. Energy relative to competing phases addresses thermodynamic stability; phonon calculations test local dynamical stability against small atomic displacements; explicit comparison of magnetic configurations tests magnetic stability; finite-temperature free energies incorporate vibrational, electronic, configurational, and, where important, magnetic contributions; and order-disorder and kinetic effects address whether the predicted state can be reached and retained experimentally. A metastable compound above the thermodynamic convex hull may still be synthesizable when transformation barriers are sufficiently large.

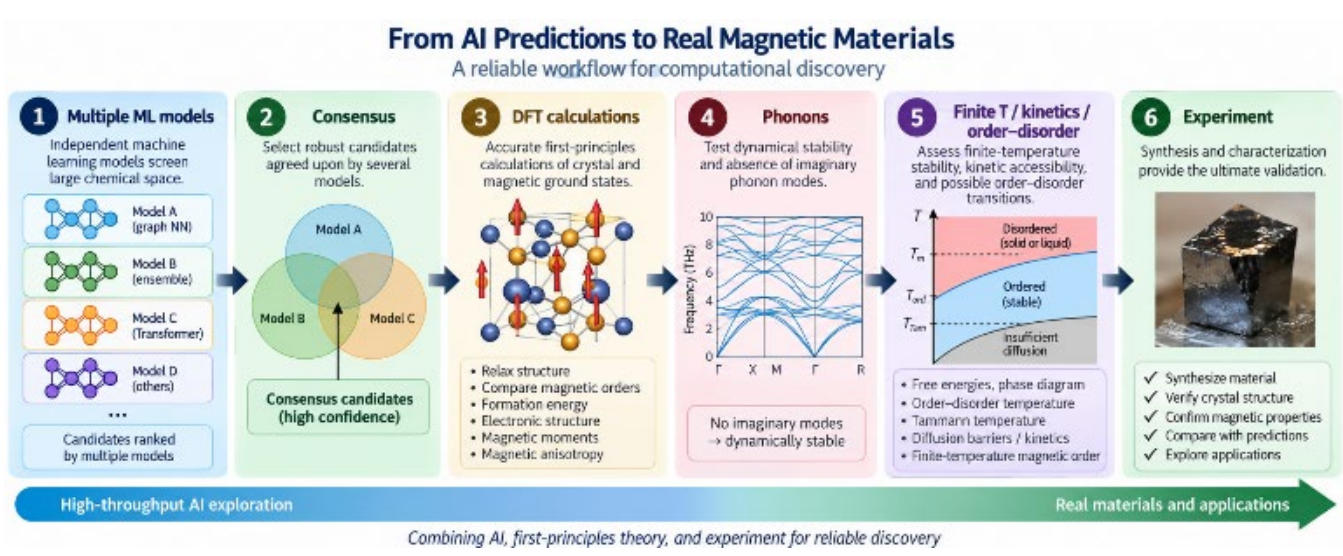

Fig.2 A proposed robust workflow for the computational prediction of new materials.

Together, these stages form a sequence of increasingly stringent but physically distinct tests. Each adds information that is generally absent from the preceding stage, and the terms stability and synthesizability should therefore not be used interchangeably: thermodynamic, dynamical, magnetic, chemical-order, and kinetic criteria address different questions. Experimental synthesis and characterization remain the final validation because neither ML nor first-principles calculations alone can establish synthesizability. The reliability of a computational discovery therefore depends less on confidence in any single model than on robustness across independent statistical, electronic-structure, magnetic, dynamical, thermodynamic, kinetic, and experimental tests.

Let us discuss an important issue that both database construction and ML/AI searches currently ignore. In recent years, many magnetic systems have been predicted as candidates for new rare-earth-free permanent magnets. Most are ordered compounds containing Fe and Co in phases previously known as disordered. For a predicted ordered compound, the relationship between the order–disorder temperature $T_{\mathrm{od}}$ and the Tammann temperature [4-5] $T_{\mathrm{Tam}}$ gives a useful first-pass criterion for whether the *ordered phase* can realistically be synthesized.

The Tammann temperature is approximately

$$T_{\mathrm{Tam}} \simeq (0.5 - 0.67)T_m,$$

where $T_m$ is the melting point, above which atomic diffusion becomes sufficiently rapid for substantial solid-state equilibration. $T_{\mathrm{od}}$ is the temperature below which the chemically ordered structure is thermodynamically favored.

The most favorable situation is therefore

$$\boxed{T_{\mathrm{Tam}} < T_{\mathrm{od}} < T_m.}$$

This creates a **kinetic–thermodynamic synthesis window**

$$\boxed{T_{\mathrm{Tam}} < T_{\mathrm{synth}} < T_{\mathrm{od}}}.$$

Within this window, atoms are mobile enough to diffuse toward equilibrium, while the ordered phase is simultaneously thermodynamically preferred. This suggests a particularly useful quantitative descriptor for computational materials screening:

$$\boxed{\Delta T_{\mathrm{synth}} = T_{\mathrm{od}} - T_{\mathrm{Tam}}.}$$

Then

$$\Delta T_{\mathrm{synth}} > 0$$

is a favorable indicator, while a larger positive $\Delta T_{\mathrm{synth}}$ generally implies greater experimental accessibility. One can also define a dimensionless **synthesis-window parameter**

$$\boxed{S = \frac{T_{\mathrm{od}} - T_{\mathrm{Tam}}}{T_m}}$$

or, if $T_m$ is unavailable,

$$S' = \frac{T_{\mathrm{od}}}{T_{\mathrm{Tam}}}.$$

Thus $S' > 1$ is favorable, $S' \sim 1$ is marginal, and $S' < 1$ indicates a kinetic obstruction to synthesis, even though the ordered compound may be on the convex hull. Commonly used synthesizability criteria based on convex-hull distance do not apply to such materials.

Thus, testing whether fabrication of a given compound requires kinetically inaccessible ordering is a useful additional filter beyond the usual energy-above-hull + phonon-stability criteria because it introduces the missing connection between thermodynamic stability and kinetic accessibility.

We now use the predicted ordered $Fe_3CoB_2$ compound [6] as a concrete test of the finite-temperature chemical-order criterion. We examine how to identify a candidate that appears favorable in a T=0 ordered-structure calculation as unsynthesizable by testing it against a chemically disordered counterpart. To quantify the Fe/Co ordering tendency, we extracted the parent $Fe_2B$ structure from the Materials Project and relaxed it in the ferromagnetic state using the Vienna Ab initio Simulation

Package (VASP) [7], with forces converged to below 0.01 eV/Å per atom. We then used ATAT [8] to generate the cluster expansion (CE) on this parent lattice.

Figure 3 displays the CE analysis of Fe/Co chemical ordering on the $Fe_2B$ parent lattice. The ordered ground-state hull lies below the random-alloy energies across the intermediate composition range, showing that Fe/Co chemical ordering is energetically favored at low temperature. The strongest ordering tendency occurs near x = 0.5, where the ground-state ordered phase is FeCoB (space group Ima2). At the $Fe_3CoB_2$ composition, x = 0.25, the lowest-energy ordered structure has orthorhombic Fmm2 (No. 42) symmetry and lies below the previously reported Cmmm (No. 65) structure [5], but its separation from the random-alloy energy is much smaller. The analytic CE predictions for the random alloy and finite-cell special quasi-random structures (SQS) [9] show the same overall trend, providing an independent consistency check on the random-alloy reference used to estimate the ordering energy.

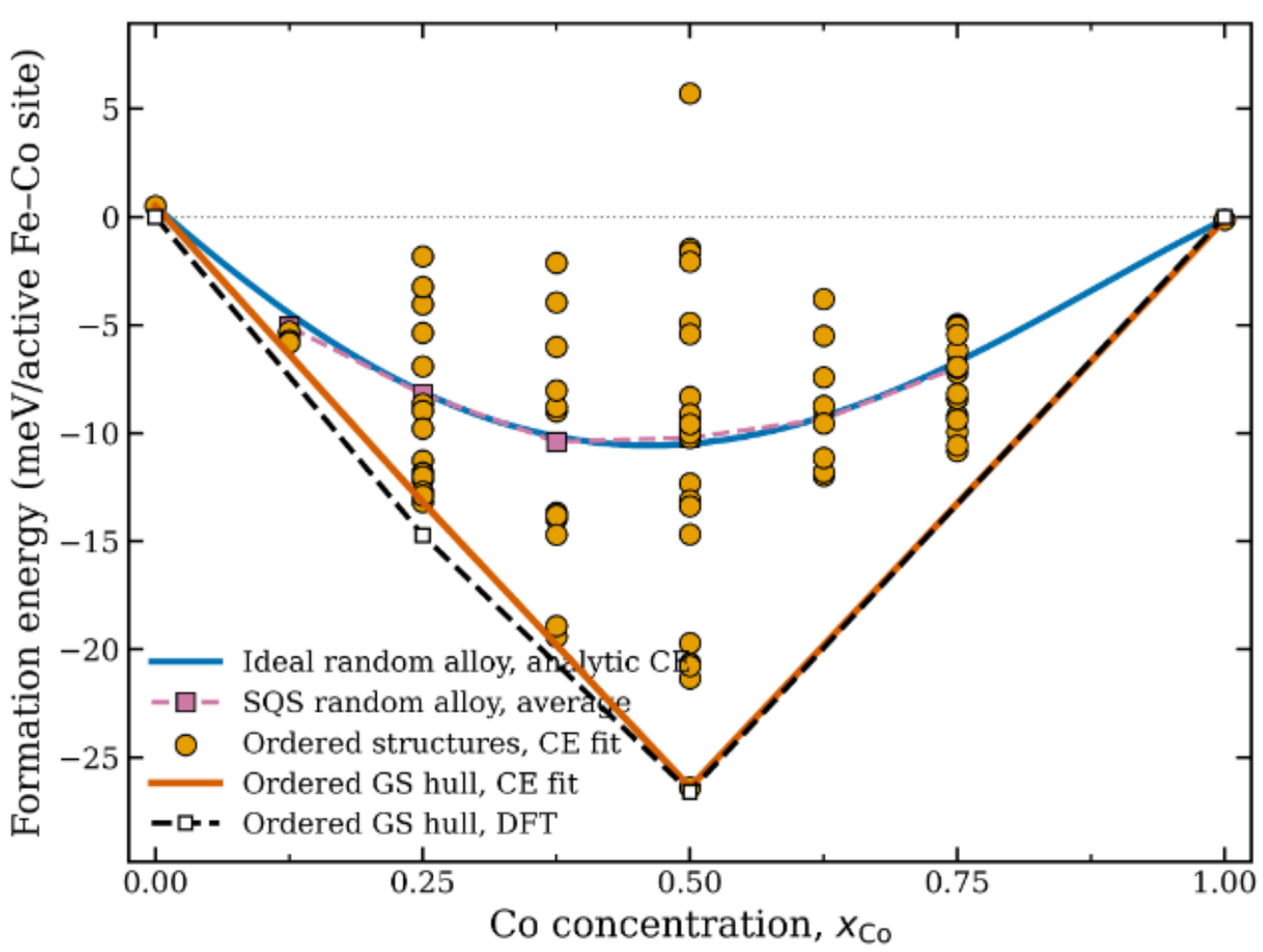

FIG. 3. Cluster-expansion analysis of Fe/Co chemical ordering in the ferromagnetically relaxed $Fe_2B$parent lattice. All energies are reported in meV per active Fe–Co atom.

Both the CE and SQS estimates agree that the ordered $Fe_3CoB_2$ configuration is lower in energy than the random Fe/Co arrangement, but by only about 5 meV per magnetic atom. This small energy scale indicates a weak driving force for long-range chemical ordering.

The mean-field chemical order–disorder temperature at x=0.25 was estimated as $T_{\mathrm{od}} = 2(E_{rand} - E_{ord})/k_B \approx 118$ K using $E_{rand}$ from CE or 116 K using $E_{rand}$ from an SQS. In a separate CE where all structures were independently relaxed in the ferromagnetic state, we obtained $T_{\mathrm{od}} \approx 174$ K. All these temperatures satisfy the above criterion $S < 0$ ( $S' < 1$ ) with a wide margin, as the melting temperature $T_{\mathrm{m}}$ in these alloys is around 1500 K.

The low chemical order-disorder temperature has a direct physical implication. The Fmm2 configuration is the T=0 chemical ground state identified at the $Fe_3CoB_2$ composition, but its low ordering energy makes it kinetically inaccessible through any annealing protocol.

## Conclusion

Reliable AI/ML discovery of magnetic materials requires physically complete data and a validation workflow that tests distinct sources of uncertainty rather than relying on database size or the accuracy of a single model. First, a database's relevant measure is its information content: increasing model complexity generally cannot recover missing structural, chemical, magnetic, or finite-temperature classes. Second, disagreement among independently built ML models can flag regions poorly constrained by the available training data. Third, validate candidate materials through explicit electronic-structure and magnetic calculations, dynamical and thermodynamic stability tests, finite-temperature and kinetic analysis, and ultimately experiment. The Fe-Co-B example shows why this hierarchy matters: an ordered structure that is energetically preferred at T=0 K can still have such a low order-disorder temperature that the predicted long-range chemical order cannot be stabilized by any heat-treatment procedure. In such cases, AI/ML is therefore most useful when coupled to magnetic and order-disorder physics and rigorous verification. In this sense, Shannon's 1956 warning in The Bandwagon remains relevant: rapid methodological progress should be accompanied by equally careful assessment of the information and physical assumptions on which the predictions depend.

## Acknowledgment

Work at Ames National Laboratory was supported by the U.S. Department of Energy (DOE), Office of Science, Basic Energy Sciences, Materials Sciences and Engineering Division. Ames National Laboratory is operated for the U.S. DOE by Iowa State University under Contract No. DE-AC02-07CH11358. Work at UNL was supported by the U.S. DOE Established Program to Stimulate Competitive Research (EPSCoR) Grant No. DE-SC0024284. Computations were performed at the High-Performance Computing facility at Iowa State University and the Holland Computing Center at the University of Nebraska.

## References

[1] C. E. Shannon, The Bandwagon, IRE Transactions on Information Theory, 2, 1, 3–11 (1956).
[2] J. Riebesell, R. E. A. Goodall, P. Benner, Y. Chiang, B. Deng, G. Ceder, M. Asta, A. A. Lee, A. Jain and K. A. Persson, A framework to evaluate machine learning crystal stability predictions, Nature Machine Intelligence, 7, 836–847 (2025)
[3] W. Xu, R. Y. Sanspeur, A. Kolluru, B. Deng, P. Harrington, S. Farrell, K. Reuter and J. R. Kitchin, Proceedings of the National Academy of Sciences, 122, 27 (2025)
[4] G. Tammann, Kristallisieren und Schmelzen, Leipzig: Johann Ambrosius Barth, 1903
[5] D. R. Gaskell, Introduction to the Thermodynamics of Materials, 5th ed. (2008)
[6] W. Xia, M. Sakurai, B. Balasubramanian, T. Liao, R. Wang, C. Zhang, H. Sun, K.-M. Ho, J. R. Chelikowsky, D. J. Sellmyer, and C.-Z. Wang, Proceedings of the National Academy of Sciences 119, e2204485119 (2022).
[7] G. Kresse and J. Furthmüller, *Phys. Rev. B* 54, 11169 (1996).
[8] A. van de Walle, M. Asta, and G. Ceder, Calphad 26, 539–553 (2002).

[9] A. Zunger, S.-H. wei, L.G. Ferria and J. E. Bernard, Physical Review Letters 65, 353 (1990)